\RequirePackage{lineno}
\documentclass[prl,aps,amsfonts,amssymb,nofootinbib,twocolumn,linenumbers,amsmath,superscriptaddress]{revtex4}
\usepackage{graphics}
\usepackage{hyperref}  
\usepackage{epsfig} 
\usepackage{color}   
  
  \usepackage{hyperref}
\hypersetup{
  colorlinks,
  citecolor=magenta,
  linkcolor=blue,
  urlcolor=blue}

\usepackage{graphicx}% Include figure files

\usepackage{bm}% bold math
\usepackage{hyperref}
\begin{document}

%\linenumbers

\title{Duality and domain wall dynamics in a twisted Kitaev chain}

\author{C. M. Morris}
\affiliation{The Institute for Quantum Matter, Department of Physics and Astronomy, The Johns Hopkins University, Baltimore, MD 21218, USA}

\author{Nisheeta Desai}

\affiliation{Department of Physics \& Astronomy, University of Kentucky, Lexington, KY 40506}

\author{J. Viirok}
\author{D. H{\"u}vonen}
\author{U. Nagel}
\author{T. R{\~o}{\~o}m}

\affiliation{National Institute of Chemical Physics and Biophysics, Akadeemia tee 23, 12618 Tallinn, Estonia}

\author{J.W. Krizan}
\affiliation{Department of Chemistry, Princeton University, Princeton, NJ 08544, USA}

\author{R. J. Cava}
\affiliation{Department of Chemistry, Princeton University, Princeton, NJ 08544, USA}

\author{T. M. McQueen}
\affiliation{The Institute for Quantum Matter, Department of Physics and Astronomy, The Johns Hopkins University, Baltimore, MD 21218, USA}
\affiliation{Department of Chemistry, The Johns Hopkins University, Baltimore, MD 21218, USA}
\affiliation{Department of Materials Science and Engineering, The Johns Hopkins University, Baltimore, MD 21218, USA}

\author{S. M. Koohpayeh}
\affiliation{The Institute for Quantum Matter, Department of Physics and Astronomy, The Johns Hopkins University, Baltimore, MD 21218, USA}
\affiliation{Department of Materials Science and Engineering, The Johns Hopkins University, Baltimore, MD 21218, USA}

\author{Ribhu K. Kaul}
\email{ribhu.kaul@uky.edu}
\affiliation{Department of Physics \& Astronomy, University of Kentucky, Lexington, KY 40506}

\author{N. P. Armitage}
\email{npa@jhu.edu}
\affiliation{The Institute for Quantum Matter, Department of Physics and Astronomy, The Johns Hopkins University, Baltimore, MD 21218, USA}

\date{\today}

\begin{abstract}
{\bf The Ising chain in transverse field is a paradigmatic model for a host of physical phenomena, including spontaneous symmetry breaking, topological defects, quantum criticality, and duality.  Although the quasi-1D ferromagnet CoNb$_2$O$_6$ has been put forward as the best material example of the transverse field Ising model, it exhibits significant deviations from ideality.  Through a combination of THz spectroscopy and theory, we show that CoNb$_2$O$_6$ in fact is well described by a different model with strong bond dependent interactions, which we dub the {\it twisted Kitaev chain}, as these interactions share a close resemblance to a one-dimensional version of the intensely studied honeycomb Kitaev model. In this model the ferromagnetic ground state of CoNb$_2$O$_6$ arises from the compromise between two distinct alternating axes rather than a single easy axis. Due to this frustration, even at zero applied field domain-wall excitations have quantum motion that is described by the celebrated Su-Schriefer-Heeger model of polyacetylene.  This leads to rich behavior as a function of field.  Despite the anomalous domain wall dynamics, close to a critical transverse field the twisted Kitaev chain enters a universal regime in the Ising universality class.   This is reflected by the observation that the excitation gap in CoNb$_2$O$_6$ in the ferromagnetic regime closes at a rate precisely twice that of the paramagnet.  This originates in the duality between domain walls and spin-flips and the topological conservation of domain wall parity.   We measure this universal ratio `2' to high accuracy -- the first direct evidence for the Kramers-Wannier duality in nature.}
\end{abstract}

\maketitle

\begin{figure*}[t]
\includegraphics[width=1.5\columnwidth]{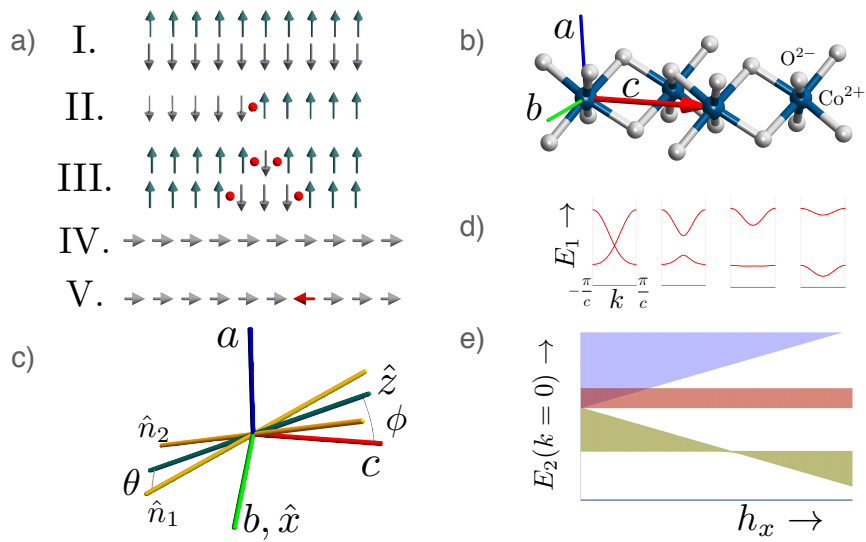}
\caption{(a) Wavefunctions for the simple transverse field Ising chain, $H= -\sum_i\left (J\tau^z_i\tau^z_{i+1}+h_x \tau^x_i\right )$ in the magnet when $J\gg h_x$ (I-III) and in the paramagnet when $J\ll h_x$ (IV,V). I. Spontaneous symmetry breaking selects one of two degenerate ground states II. Non-local single domain wall elementary excitation III.  A single spin flip fractionalizes into at least two domain wall excitations  IV. The quantum paramagnet at high transverse field with a singly degenerate ground state. V. A single spin-flip is the elementary excitation of the paramagnet.  (b) Crystal structure~\cite{Heid1995} of CoNb$_2$O$_6$ showing the zigzag chain of Co$^{+2}$ ions and the edge sharing network of distorted O$^{2-}$ octahedra.  Crystallographic $a$, $b$, $c$ directions are indicated. (c) In yellow are indicated the two alternating Ising directions $\hat n_1$ and $\hat n_2$, which by crystal symmetry are related by $\pi$-rotation about the $b$ axis. The $\hat x$ spin direction is identified with the $b$-axis. The $\hat z$ spin direction (which lies in $a-c$ plane) also lies in the plane of the Ising axes and bisects the angle between $\hat n_1$ and $\hat n_2$ and is the direction of polarization of the ferromagnetic moment of the twisted Kitaev chain. From previous experimental studies~\cite{Heid1995,Kobayashi2000}, we infer that $\hat z$ makes an angle $\phi\approx\pm 31^\circ$ with the $c-$axis. From our analysis here we conclude the angle $\theta$ that $\hat n_1,\hat n_2$ makes with the $\hat z$-axis is approximately 17$^\circ$.  This fixes the Ising directions with respect to the crystal axes.   (d) The dispersion of the single domain wall band is split into two in a fashion similar to the Su-Schrieffer-Heeger model.  The bands evolve with increasing $h_x$. (e) Cartoon of the evolution with $h_x$ of the continua of two domain wall kinetic energies with total momentum, $k_{\rm tot}=0$. The continua have three branches, when the two domain walls are both in the upper band (blue), both in the lower band (green) or when they are in different bands (red). }
\label{fig:intro}
\end{figure*}

The transverse field Ising model (TFIM) describes a 1D system of spin-1/2 moments with interactions that favor spins aligned along an easy axis~\cite{Onsager44a,KramersWannier,Sachdev2011,Mussardo10a,Pfeuty1970,McCoy1978,Zamolodchikov1989}. At small transverse fields the ground state is two-fold degenerate (Fig.\ref{fig:intro}(a)I); the system magnetizes by picking a ground state by the phenomena of spontaneous symmetry breaking. The elementary excitations of the 1D Ising magnet are domain walls, which form between the two ground states (Fig.~\ref{fig:intro}(a)II).  These are topological since their parity cannot be altered by the action of any local operator. Experimental probes like THz can only create an even number of domain walls (Fig.~\ref{fig:intro}(a)III).   An applied transverse field increases quantum fluctuations, which results in domain wall motion.   At a quantum critical point (QCP), there is a transition to a singly degenerate paramagnetic ``quantum disordered" state (Fig.~\ref{fig:intro}(a)IV) where the elementary excitations are conventional ``spin-flip'' quasi-particles that can be created and destroyed individually (Fig.~\ref{fig:intro}(a)V). The QCP is in the much studied 1+1 D Ising universality class~\cite{Sachdev2011}.  In ground breaking work, Kramers and Wannier~\cite{KramersWannier} showed in 1941 that the single domain walls (Fig.~\ref{fig:intro}(a)II) and spin-flip quasiparticles (Fig.~\ref{fig:intro}(a)V) can be related to one another by a so-called duality transformation, so that even a dense ensemble of domain walls can be transformed exactly to spin-flips near the QCP. This concept of duality has had a profound impact on disparate branches of physics, from high energy and string theory to condensed matter~\cite{polyakov1987:gauge,polchinski,Fisher2004}.

Despite the extensive theoretical impact of the Ising chain, it has had few material realizations.  Coldea {\it et al.}~\cite{Coldea2010} pointed out that CoNb$_2$O$_6$ has features that could make it an excellent example. Magnetic Co$^{+2}$ ions, which tend to have strong uniaxial anisotropy are arranged in chains that magnetize at low temperature with an easy axis in the $a$-$c$ plane.  These ferromagnetic chains are arranged antiferromagnetically in an order that is long-range commensurate at the lowest temperatures.  A magnetic field along the $b$-axis is transverse to the moments and a field $B_c\approx 5.3$ destroys the magnetism at a QCP (See SI for further discussion on phase diagram). A number of quantitative observations strengthened the TFIM picture.  At $B_b=0$ in the low-$T$ commensurate ordered state an extremely rich spectrum of nine bound states described by Airy function solutions to the 1D Schr\"odinger equation with a linearly confining potential~\cite{Coldea2010,Morris14a} were observed.  Close to the QCP, scaling of NMR response consistent with the 1D Ising universality class~\cite{imai2014,steinberg2019nmr,ongcv} and a neutron spectral function consistent with an integrable field theory~\cite{Coldea2010} was found. While it appears that CoNb$_2$O$_6$ is an excellent realization of Ising quantum {\it criticality},  significant deviations from the TFIM have also been noted~\cite{Coldea2010,kjall2011:e8,robinson2015:break,fava2020glide,Heid1995,Kobayashi2000}.  Here we establish that this material is in fact described by a different model of fundamental interest, the twisted Kitaev chain~\cite{kitaev2006anyons}.

\begin{figure*}[t]
\includegraphics[trim={0cm 4cm 0cm 4cm},clip,width=2\columnwidth]{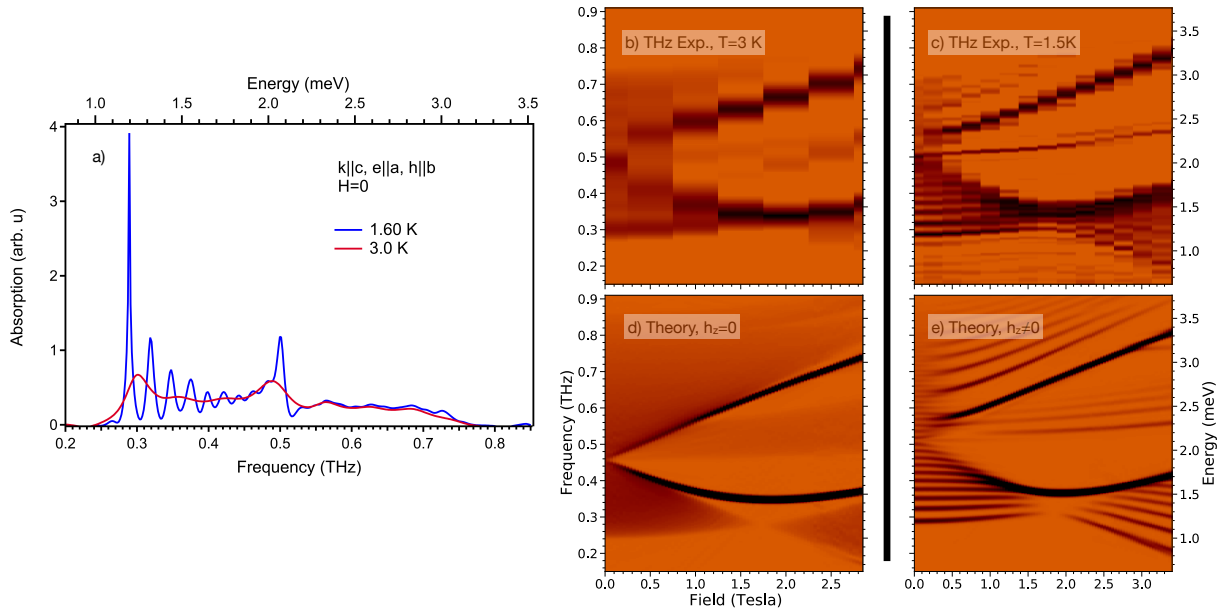}
\caption{ TDTS data on CoNb$_2$O$_6$ and theoretical simulations of the twisted Kitaev chain spectral function.  a) TDTS absorption spectra at zero transverse field at temperatures above and below the ordering temperature. (b-c) TDTS absorption shown as a function of a $b$-axis field $B_b$ and the frequency of light.  (b) The data taken at $T=3$K is just above the critical temperature. In a finite $B_b$ the continua at zero field splits into three broad features with distinct characteristic energy and field dependence.  (c) Data taken below the critical temperature at $T=1.5$K. (d-e) Theoretical calculations of the ground state spectral function $A(\omega)\equiv \sum_\alpha |\langle \alpha | \sum_i \tau^x_i |0\rangle|^2 \delta(\omega-E_\alpha)$ using the density matrix renormalization group method, which provides a map of the $k=0$ excitation spectrum $E_\alpha$ for the twisted Kitaev chain model. $A(\omega)$ with (d) $h_z=0$  and  (e) with $h_z=0.02$ meV.}
\label{FieldEvolutions}
\end{figure*}

In magnetic insulators, absorption spectra using time-domain THz spectroscopy (TDTS) and Fourier transform infrared spectroscopy (FTIR) can be used to do extremely sensitive measurements of the quantity $\omega \chi (q = 0, \omega)$, where  $\chi $ is the zero momentum complex magnetic susceptibility (See SI).  In Fig.~\ref{FieldEvolutions}a we show TDTS absorption spectra at zero applied field.   Above the ordering transition at 3K, there are two distinct spectral regions (0.28 - 0.52 THz, 0.52 - 0.75 THz) which can be attributed to the domain wall excitations.  In previous work the higher energy  was assigned to a four domain wall excitation, but we will see here that both features result from the novel dynamics of two domain wall excitations in the twisted Kitaev chain model. Below the ordering temperature the appearance of sharp peaks can be understood with the assumption that the effect of the ordering on the 1D chain is as a longitudinal Weiss field that confines domain walls~\cite{Coldea2010,Morris14a}.   In Fig.~\ref{FieldEvolutions}b and c, we show the absorption plotted as a function of transverse field at temperatures right above the transition ($T=3$K) and at ($T=1.5K$) in the ordered state.  This data was taken with B$_{THz} \parallel$ b, so it is most sensitive to the ferromagnet with the spins in the $a-c$ plane.

Most striking in this data are a number of qualitative deviations from the simplest TFIM expectations. First, as also apparent in Fig.~\ref{FieldEvolutions}a, the two domain wall excited states have a broad dispersion even at $B_b=0$. Since this dispersion exists both above and below the ordering temperature, it cannot be generated by mean-field interchain interactions and must be intrinsic to the chain.  This effect is absent in the TFIM at zero transverse field, where the domain walls would be motionless.  The origin of the effective transverse field has been unclear since early work~\cite{Coldea2010,kjall2011:e8,Morris14a}, but has recently been attributed to terms beyond the Ising interaction arising in a symmetry analysis~\cite{fava2020glide}. Since the dispersion scale is comparable to the gap it is plausible that it has the same origin as the gap itself, a point we return to below. Second, and most strikingly, with applied field the spectrum splits into three features. The lowest energy feature has a threshold for excitations that depends non-monotonically on $B_b$ and one can see that its bandwidth goes through a minimum at an exceptional field near 2 T.  The middle feature depends weakly on $B_b$ at low fields. Finally the high energy feature does not decrease in energy (as it would for a 4 domain wall excitation), but surprisingly {\it increases} with increasing field.  In the low-$T$ ordered phase, the bound states fall within the energies of the higher temperature two particle continua and show the same narrowing of the band width around 2 T, before increasing again.  The upper energy region even exhibits discrete peaks which are reminiscent of the lower energy bound states.  All of these aspects are in stark contrast to the TFIM in which at $B_b=0$ the excitations have no dispersion, acquiring one that scales with the applied field with a monotonically decreasing gap that goes to zero at the QCP.  In the present case, only beyond a substantial field of $B_b\approx 2T$ does one find the kind of gap closing expected from the TFIM, and only for the low energy excitations does one find the kind of gap closing expected from the TFIM.  These observations indicate that even at a qualitative level CoNb$_2$O$_6$ is not described by a simple TFIM.

We now turn to a model for these remarkable observations. In CoNb$_2$O$_6$, Co$^{2+}$ ions are surrounded by an edge sharing network of distorted octahedra of oxygen ions forming relatively isolated one-dimensional chains. Due to the zig-zag structure, $c$-axis crystal translations connect a Co ion to its next-nearest neighbors along the chain (see Fig.~\ref{fig:intro}(b)). It has been argued recently that with the high symmetry of perfect edge sharing octahedra, such 3d$^7$ Co ions have a spin-orbit coupled pseusdospin-1/2 ground state and an Ising interaction with an easy-axis that is bond-dependent (perpendicular to the local Co-O exchange planes)~\cite{khalliulin2018:cod7,sano2018kitaev}. 
Inspired by these studies, we assume that the Co-Co nearest neighbor interactions are restricted to the Ising form, but due to the complex local quantum chemistry from the octahedral distortions we do not constrain the directions theoretically, rather we determine them by symmetry and experimental input. Our model is simplified by the observation that once the Ising axis for one bond is fixed {\em the relative orientations of all others Ising easy axes in the crystal are constrained by the Pbcn space group}. For example it follows from the $c$-translation that the Ising directions must alternate along the chain,
\begin{equation}
  \label{eq:twisted}
\mathcal{H}_K = - K \sum_{i\in e} \left ( \tau_i^{\hat n_1} \tau_{i+1}^{\hat n_1} +
  \tau_{i+1}^{\hat n_2} \tau_{i+2}^{\hat n_2}\right ),
\end{equation}
where $ \tau_i^{\hat n}=\hat n \cdot \vec \tau_i $ is the Pauli operator on the $i^{\rm th}$ site along the chain projected along the $\hat n$ `Ising' direction and the sum is taken only on the even sites.  We call this Hamiltonian the {\it twisted Kitaev chain}.  It was introduced previously as a way to interpolate between the Ising chain, $\hat n_1 \cdot \hat n_2 =1$ and the Kitaev model, $\hat n_1 \cdot \hat n_2 =0$~\cite{you2014:qcm}. The space group Pbcn of CoNb$_2$O$_6$ includes ${\mathcal R}^\pi_b$, a $\pi$-rotation along the $b$-axis passing through a Co site. This operation interchanges $\hat n_1, \hat n_2 $ leading to the conclusion that they are related by a $\pi$ rotation about the $b$-axis, Fig.~\ref{fig:intro}(c). Thus, the Ising axes throughout the crystal are determined by two angles $\phi$ (the angle that the chain-magnetization makes with the $c-$axis) and $2\theta$ (the angle between $\hat n_1$ and $\hat n_2$).  We infer based on past work that $\phi\approx 31^\circ$~\cite{Kobayashi2000}, and as we shall find below $\theta\approx 17^\circ$.  Generically (away from $\hat n_1 \cdot \hat n_2 \neq 0$), the frustration that arises from the non-collinearity of the Ising axes is released by forming a ferromagnet that is polarized in a direction that makes the smallest equal angle between each of $\hat n_1$ and $\hat n_2$~\cite{you2014:qcm}\footnote{An exception is when $\hat n_1 \cdot \hat n_2=0$.  Then the model reduces to a one-dimensional section of the honeycomb Kitaev model for which the ground state is pathologically degenerate.  Here we are interested in the more generic behavior.}, which spontaneously breaks the ${\mathcal R}^\pi_b$ symmetry. We argue that it is this mechanism that is responsible for the ferromagnetism in CoNb$_2$O$_6$. As we shall see the consequence of this frustration is extremely rich domain wall dynamics, which quantitatively describe the THz observations.  It is instructive to rewrite ${\mathcal H}_K$ by introducing global $\hat x,\hat y,\hat z$ axes for the spin operators. We will identify $\hat x$ with the crystal $b$-axis and $\hat z$ with the axis orthogonal to $\hat x$ that is also contained in the plane spanned by $\hat n_1$ and $\hat n_2$ (Fig.~\ref{fig:intro}). In these axes one has,
\begin{eqnarray}
  \label{eq:hspinxz}
 \mathcal{H} &=& -K \sum_i  \Big[ \cos^2(\theta) \tau^z_i \tau^z_{i+1}+\sin^2  (\theta) \tau^x_i \tau^x_{i+1}   \nonumber \\
  &+&  \frac{\sin(2\theta)}{2}(-1)^i ( \tau^x_i \tau^z_{i+1}+ \tau^z_i \tau^x_{i+1})    \Big] \nonumber \\
  &-& h_x \sum_i \tau_i^x - h_z \sum_i \tau_i^z.
\end{eqnarray}
We have included magnetic fields along the $\hat x$ and $\hat z$ directions. Since the $\hat x$ and $b$ directions are coincident, $h_x$ is simply the externally applied transverse field, so  $h_x=g \mu_B B_b/2$. We will be interested in the regime when $\theta < 45^\circ$ and so the ferromagnet is polarized in the $\hat z$ direction\footnote{Although not central to our discussion here, there are two families of chains that are in all respects identical except they are rotated about the c-axis by $\pi$ with respect to each other. These two families will thus have a rotated pair of $\hat n$ easy axes and ferromagnetic polarization. This implies theoretically that two families of chains will have the same $\theta$ but will have $\pm$ values of $\phi$, precisely as observed in early experiments~\cite{Kobayashi2000}.  } In the ordered state, an effective $h_z$ term is created by the Weiss mean field of the nearest neighboring chains that are polarized in the $-z$ direction.
The oscillating term which has been discussed in another recent work~\cite{fava2020glide} is allowed because the $c$-translation symmetry encompasses two spins. The Ising operation ${\mathcal R}^\pi_b$ sends lattice site $i\rightarrow -i$ and rotates the spins about $\hat x$, and as expected is a symmetry only so long as $h_z=0$.  

To get intuition for the physics of $\mathcal{H}_K$, we consider it in the limit $\theta, h_x/K, h_z/K \ll 1$, using degenerate perturbation theory. The unperturbed states are classified by the number of domain walls present. The largest contribution for the THz response is from the two domain wall sector which is generated by single spin flips. The projection of ${\mathcal H}$ into this degenerate two domain wall sector results in three terms: $\mathcal{H}_d$ the kinetic energy of single domain walls (due to $K$ and $h_x$), $\mathcal{H}_c$ a linear confining potential between domain wall pairs (due to $h_z$), and $\mathcal{H}_l$ which is a short range interactions between the domain walls when they are a single lattice spacing apart.  The kinetic energy of a domain wall is described by,
\begin{eqnarray}
  \label{eq:ssh}
\mathcal{H}_{d} &=& - \sum_n \left [\left ( h_x + (-1)^n K \sin (2\theta) \right ) (d^\dagger_nd_{n+1}+d^\dagger_{n+1}d_n )\right.\nonumber\\
&+&\left. K \sin^2(\theta)  (d^\dagger_nd_{n+2}+d^\dagger_{n+2}d_n ) \right ],
\end{eqnarray}
where $d_n$ destroys a domain wall at a dual site $n$.

Interestingly to leading order in the perturbations (the first term), ${\mathcal H}_d$ has precisely the form of the Su-Schrieffer-Heeger model~\cite{ssh1988rmp} with a modulated strong-weak hopping. In this analogy the single domain walls in CoNb$_2$O$_6$ map to the electrons in polyacetylene. While in polyacetylene the electronic motion is fixed by chemistry, here we can tune the band structure of domain walls by varying $h_x$, shown in Fig.~\ref{fig:intro}(d). At $h_x=0$ the hopping alternates sign between bonds which results in the single domain wall dispersion relation whose minimum is shifted by $\pi/c$ from that of the TFIM.  As a THz experiment creates two domain walls with  $k_{\rm tot}=0$, the accessible kinetic energies of these excitations exhibit three branches depending on whether the two domain walls are in the lower band, in separate bands or both in the upper band, Fig.~\ref{fig:intro}(e). These three $k_{\rm tot}=0$ two domain wall kinetic energy branches can be identified with three prominent features in our experimental data.  When a longitudinal field is switched on, the linear confining potential $H_c$ breaks the continua into discrete bound states, precisely as observed at low temperature. As shown in Fig.~\ref{fig:intro}d, a simple explanation arises for the bandwidth narrowing that is most prominent at 2T. It corresponds to the nearly complete localization of the lower domain wall band! %In the polyacetylene analogy, this corresponds to the extreme limit  when electrons are localized to a pair of carbon sites, a limit often discussed pedagogically in that context~\cite{ssh1988rmp}, but a regime that can be realized in our experiment.

In CoNb$_2$O$_6$, $\theta$ is {\it not} small and hence the two and higher domain wall sectors overlap, requiring a full solution of $\mathcal{H}_K$ for quantitative reliability. This is a hard many-body problem but one that can be simulated using density matrix renormalization group and matrix product state methods~\cite{itensor}. In Fig.~\ref{FieldEvolutions}d,e we present the result of such calculations. We use $K=0.57 $ meV, $g=3.5$ and $\theta =17^\circ$.  A standard Weiss mean field approximation for the interchain couplings has been made in which $h_z=0$ models the behavior of the system just above the transition ($T=3$K) and $h_z\neq 0$ captures the effect of the interchain coupling below the transition ($T=1.5$K).  The correspondence to experiment is remarkable, validating our simple model\footnote{Our model captures the THz data which studies the long distance $k=0$ response. A full description of the short distance physics at large momenta might require further neighbor exchanges~\cite{Coldea2010}}.  We note that the splitting when the domain walls are both in the lower band is smaller than when the domain walls are in the higher band, which can be traced back to the difference in the effective masses that the domain walls experience in the lower and upper bands, Fig.~\ref{fig:intro}d. There is likely a small transverse Weiss field from the next nearest neighbor chains,
%(an effect neglected in our calculations)
 which explains some of the minor differences.

\begin{figure}[tb]
\includegraphics[width=0.82\columnwidth]{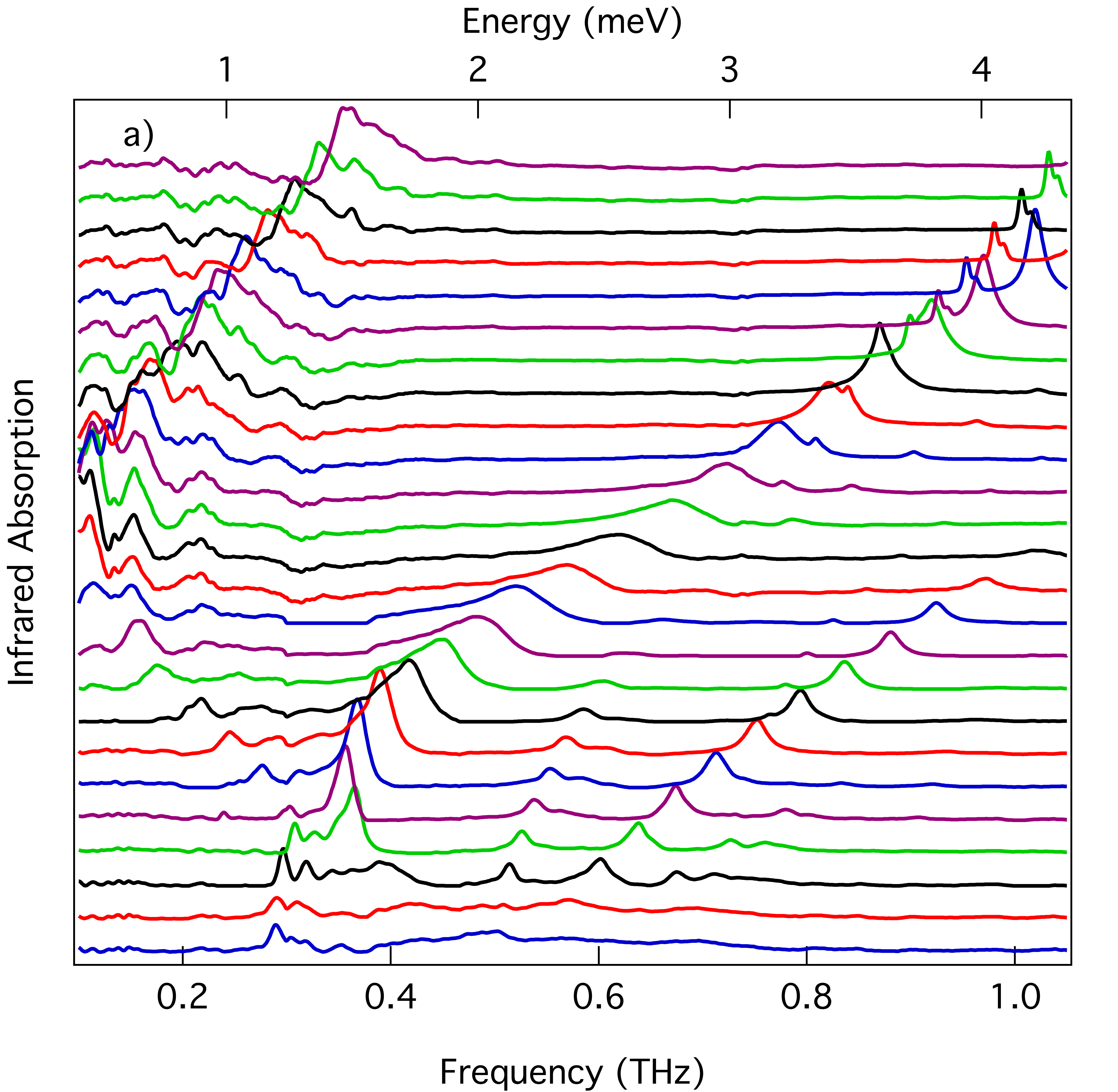}\\
\hspace{0.3cm}
\includegraphics[width=0.9\columnwidth]{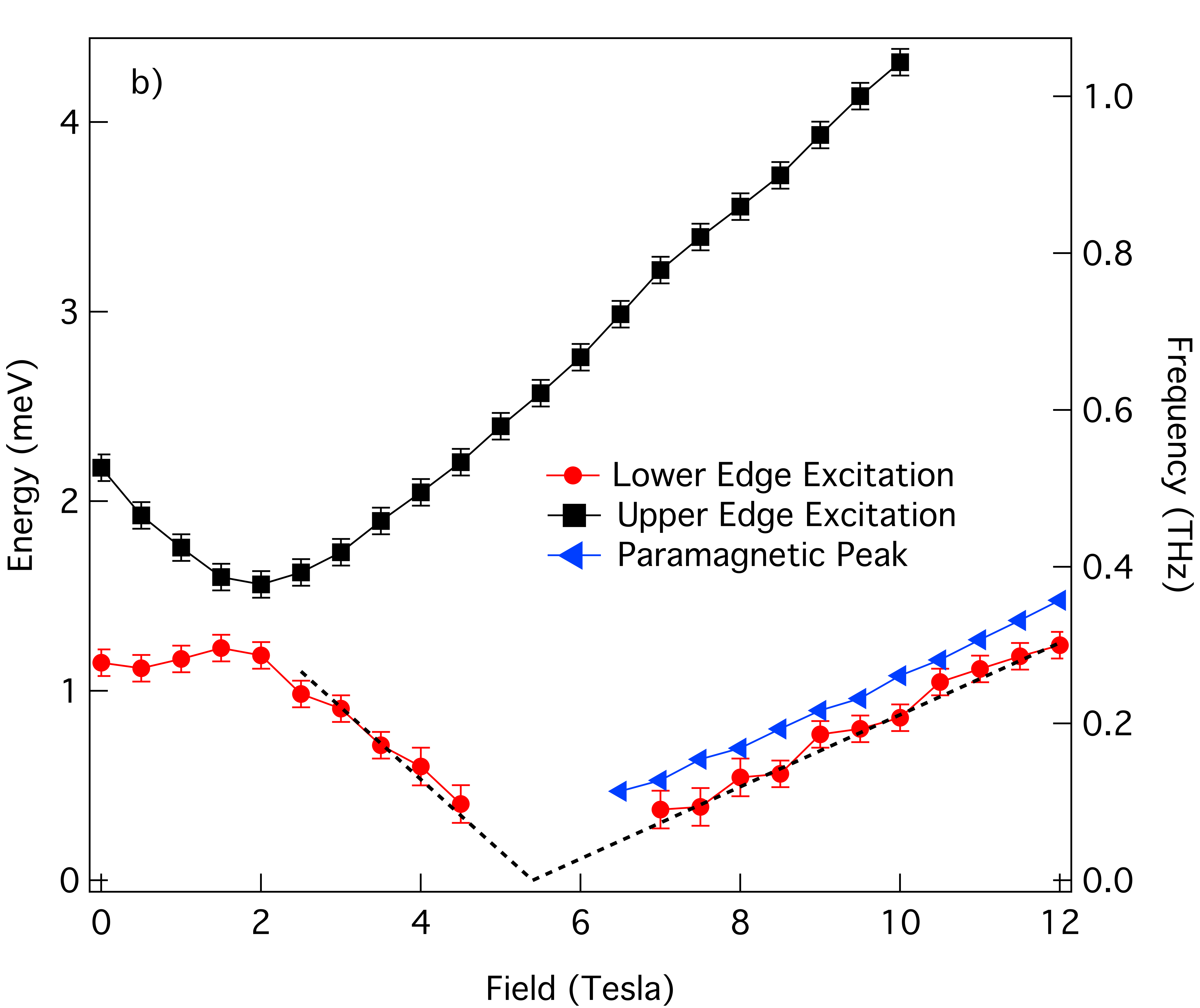}
\caption{a) FTIR absorption spectra at 2.5 K tuned with transverse field of 0 to 12 T in 0.5 T steps from bottom to top.  Spectra vertically offset.   Fine structure at the lowest frequencies arise from multiple reflections in the FTIR setup.   They do not change with temperature or frequency.  b) Plot of absorption energies as a function of transverse Field from the 2.5 K absorption spectra.   Plotted is both the lower and upper edges of the `lower energy' feature, as well as the peak energy in the paramagnetic regime.   Dashed lines are guides to the eye with a factor of 2 difference in slope. }
\label{Fig4}
\end{figure}

To understand the field evolution up to and into the paramagnetic phase, complementary experiments were performed with FTIR spectroscopy up to 12 Tesla.   Here we used unpolarized light propagating in the $b$ direction, which gives excellent intensity for all experimental fields.  In Fig.~\ref{Fig4}a one can see quite clearly the initial narrowing of the lower energy continuum, coming to a minimum band width near 2 T.  Then the lowest energy edge of the continuum softens dramatically towards zero near the known critical point near 5.3 T.   In the paramagnetic regime a single peak appears from low energy and moves to higher energy with increasing field.  This opening and closing of the gap is direct spectroscopic evidence of the quantum phase transition.  In \ref{Fig4}b we plot the energy of lower and upper edge of the excitations as well as the peak energy in the paramagnetic regime as a function of transverse field.  As $h_x$ is increased the lower edge of the domain wall continuum first increases but then goes to zero, signaling an instability to domain wall condensation.

Since the phase transition in our twisted Kitaev chain model is driven by proliferation of domain walls in the ferromagnetic regime, by arguments of universality we expect the critical regime to be described by an Ising field theory. The gap to create elementary excitations will vanish as $\Delta_\pm\sim {\mathcal A}_\pm |t|^{z\nu }$, where $\pm$ refers to the two sides of the transition, with z=$\nu$=1 ($t\equiv (B_b-B^c_b)/B^c_b$).  In the ferromagnetic regime this traces out the edge of the continuum, whereas in the paramagnetic regime it is expected to follow the quasiparticle excitation. The dashed lines in Fig.~\ref{Fig4} show that the gap goes to zero linearly on both sides of the transition consistent with Ising scaling.

We can make another striking observation.  Even though the amplitudes ${\mathcal A}_\pm$ are non-universal, the amplitude ratio ${\mathcal D}= {\mathcal A}_-/{\mathcal A}_+$, is universal and takes the value of $2$ for the 1+1 dimensional Ising model. This follows from the profound physical argument that even close to the critical point where the system is strongly interacting, the Kramers-Wannier duality~\cite{KramersWannier} relates the system at $t$ and $-t$ so that gap to create a domain wall is the same on one side of the transition as the gap to create a spin-flip particle on the other side. But because of the topological conservation law of domain wall parity, the experimental probe can only create a pair of domain walls with twice the energy and hence by duality ${\mathcal D}=2$.  Using a critical field of 5.4 T, we find from our data that this ratio is 1.9 $\pm$ 0.1, which is close to `2' as predicted.  The measurement of this universal number is the first direct evidence for the Kramers-Wannier duality and the topological conservation of domain wall parity.

We have demonstrated that while CoNb$_2$O$_6$ is an excellent realization of Ising quantum criticality, microscopically it deviates significantly from the TFIM. Instead, we have found that the rich domain wall physics uncovered by THz spectroscopy is described by another model of fundamental interest, the twisted Kitaev chain in a transverse field. Our model is inspired by a novel spin-orbit coupled superexchange theory that leads to Kitaev-like bond dependent interactions and demonstrates that Co$^{2+}$ based magnets~\cite{zhong2020weak,vivanco2020competing} may be a rich avenue for the exploration of quantum spin liquid phases.
 
Work at JHU and Princeton was supported as part of the Institute for Quantum Matter, an EFRC funded by the DOE BES under DE-SC0019331.  Work at the UK was supported by NSF DMR-1611161.  The work at NICPB was supported by institutional research funding IUT23-3 of the Estonian Ministry of Education and Research, and by European Regional Development Fund Project No. TK134.

\bibliography{QuantumMag}

\newpage

\Large {\bf Supplemental Material:  Duality and domain wall dynamics in a twisted Kitaev chain}

\normalsize

\medskip

\section{Experimental details}

\subsection{Low frequency optical measurements}

\begin{figure}[b]
\includegraphics[width=1\columnwidth]{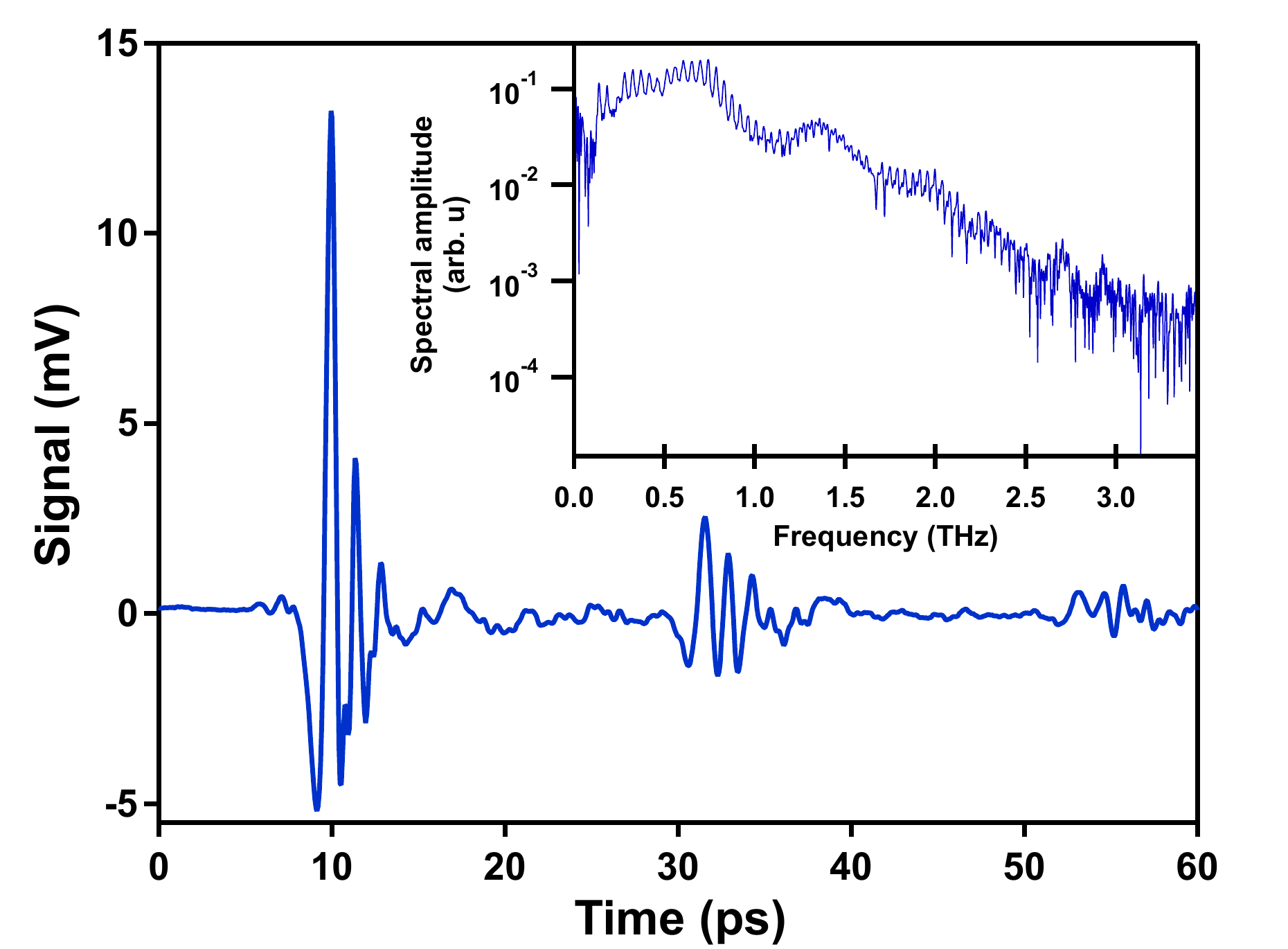}
\caption{Time-domain THz traces.  Voltage is the detected voltage in the instrumentation, but it is proportional to the electric field.  Multiple echos from the multiple reflections are seen.Inset is the Fourier tranform of the given time trace.   The fine structure is the Fabry-Perot peaks from standing waves in the sample.  }
\label{TimeTraces}
\end{figure}

Time-domain THz spectroscopy (TDTS)~\cite{RevModPhys.83.543,Nuss1998} at JHU was performed using a home-built transmission mode spectrometer that can access the electrodynamic response between 100 GHz and 2 THz (0.41 meV -- 8.27 meV) in dc magnetic fields up to 7 Tesla.  In TDTS, one excites emitter and receiver photoconductive switches sequentially with an ultrafast laser to create and detect pulses of THz radiation.  Taking the ratio of the transmission through a sample to that of a reference aperture yields the complex transmission coefficient $T(\omega)$.  If the material is an magnetic insulator with no prominent electronic low energy degrees of freedom then the resulting transmission is a function of the system's complex magnetic susceptibility as $-\ln(T(\omega)) \propto \omega \chi(q=0,\omega)$.   The time-varying magnetic field of the pulse couples to magnetic dipole excitations in the system.  Note that as the wavelength of THz range radiation is much greater than typical lattice constants (1 THz $\sim$ 300 $\mu$m in vacuum), TDTS measures the $q \rightarrow 0$ response.  The technique allows very sensitive measurements of the magnetic susceptibility at $q \rightarrow 0$ and allows the observation of small fine features that may be not apparent with other spectroscopic techniques.  The dc magnetic field was applied in the $b$ direction and measurements were performed in the Faraday geometry with the THz propagation direction parallel to the dc applied field.   Particular attention was paid to ensure alignment of the dc magnetic field with respect to this axis.   The large moment on the Co$^{+2}$ ion can give large torques for fields in the $b$ direction.   These problems were ameliorated by a particularly rigid sample holder, however Hall sensors were installed to monitor for any alignment changes due to sample torque.

In order to achieve sufficiently high resolution to discern the very sharp low energy bound states, the time domain signal is collected for 60 ps.  However, as can be seen in Fig.~S\ref{TimeTraces} this long collection time introduces difficulties, as the spectroscopy is being performed on a 600 $\mu$m thick single crystal of CoNb$_2$O$_6$ with an index of refraction of $\sim5.4$. With these parameters, reflections of the THz pulse are observed in the time domain spaced by approximately 22 picoseconds.  These reflected pulses in the time domain manifest themselves as Fabry-Perot oscillations in the frequency spectrum, with an amplitude large enough to obscure the fine magnetic structure. In principle sufficient knowledge of the frequency dependent index of refraction should make it possible to numerically remove these oscillations, however, in practice the required level of precision needed in the index of refraction is too high. Instead, we have developed a method of data analysis that exploits the fact that the index of refraction of the material changes fairly slowly with temperature (aside from the absorption due to magnetic excitations) at low temperature, and therefore referencing between scans with close temperatures (but above and below the ordering temperature where bound states develop) can remove the oscillations.  Our goal is to resolve the sharp magnetic excitations in the system at a temperature below the commensurate antiferromagnetic interchain ordering temperature $\approx$ 1.95 K~\cite{Kobayashi2000,Heid1995}. Therefore the principal step in the analysis is to divide the high resolution spectrum at low temperature by a spectrum at a higher temperature that does not have the sharp excitations, to remove the Fabry-Perot oscillations.  The method is discussed in detail in the Supplemental to Ref.~\cite{Morris14a}.

In addition to the TDTS, Fourier Transform Infrared (FTIR) experiments were performed at the National Institute of Chemical Physics and Biophysics in Tallin, with a Sciencetech SPS200 Martin-Puplett type spectrometer with a 0.3 K bolometer.  Transmission measurements were carried out at 2.5 K in a cryostat equipped with a superconducting magnet.  Fields up to 12 T in the Faraday geometry were applied.   Although a small wedge was polished into these samples they still show standing wave interferences at low frequency that must be accounted for in the analysis.

\subsection{Sample preparation}

Stoichiometric amounts of Co$_3$O$_4$ and Nb$_2$O$_5$ were thoroughly ground together by placing them in an automatic grinder for 20 minutes. Pellets of the material were pressed and heated at 950$^\circ$ C with one intermittent grinding. The powder was then packed, sealed into a rubber tube evacuated using a vacuum pump, and formed into rods (typically 6 mm in diameter and 70 mm long) using a hydraulic press under an isostatic pressure of $7$x$10^7$ Pa. After removal from the rubber tube, the rods were sintered in a box furnace at 1375$^\circ$ C for 8 hours in air.

Single crystals approximately 5 mm in diameter and 30 mm in length were grown in a four-mirror optical floating zone furnace at Johns Hopkins (Crystal System Inc. FZ-T-4000-H-VII-VPO-PC) equipped with four 1-kW halogen lamps as the heating source. Growths were carried out under 2 bar O$_2$-Ar (50/50) atmosphere with a flow rate of 50 mL/min, and a zoning rate of 2.5 mm/h, with rotation rates of 20 rpm for the growing crystal and 10 rpm for the feed rod. Measurements were carried out on oriented samples cut directly from the crystals using a diamond wheel.

Cut samples were polished to a finish of 3 $\mu$m and total thickness of $\sim600$ $\mu$m using diamond polishing paper and a specialized sample holder to ensure that plane parallel faces were achieved for the THz measurement. The discs were approximately 5 mm in diameter. 

\begin{figure}[t]
\includegraphics[width=0.85\columnwidth]{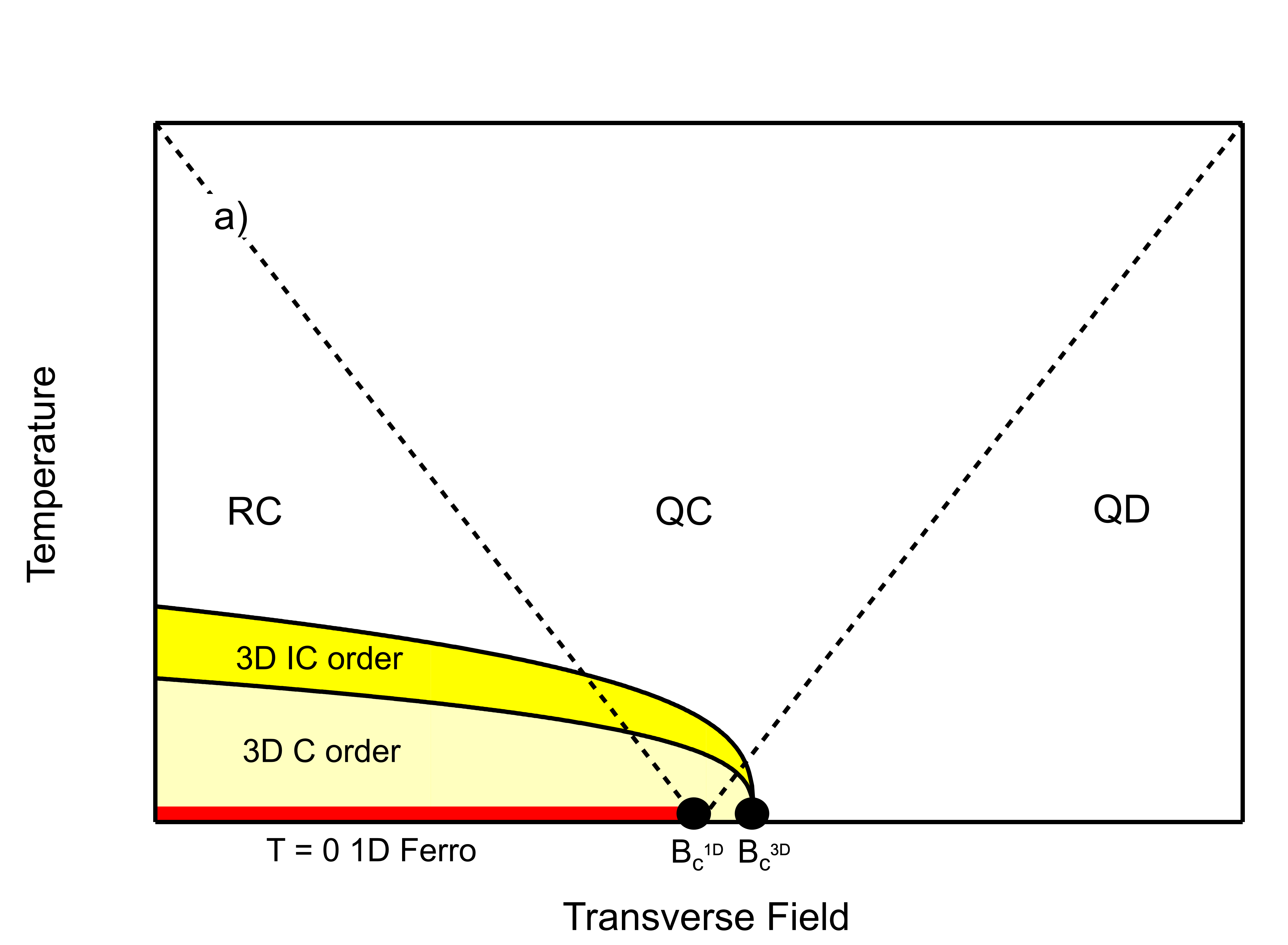} 
\caption{ a) Phase diagram expected for CoNb$_2$O$_6$ as a function of transverse magnetic field.  Labels correspond to different regimes of order or expected behavior.  IC= Incommensurate order, C = Commensurate order, RC = renormalized classical, QC = quantum critical, QD = quantum disordered.   In an isolated 1D chain true long range order is only found at T=0.   }
\label{PhaseDiagram}
\end{figure}

\section{{The phase diagram of C\lowercase{o}N\lowercase{b}$_2$O$_6$}}

A rough schematic of the phase diagram of CoNb$_2$O$_6$ as a function of transverse field is shown in Fig.~S\ref{PhaseDiagram}. At isolated ferromagnetic 1D chain is expected to have ferromagnetic order only at T=0, although there should be substantial buildup of ferromagnetic correlations at finite temperature~\cite{Hanawa1994}.  Couplings to other chains cause CoNb$_2$O$_6$ to have an ordering at approximately 3K to a 3D incommensurate ordered state and then at $\approx$1.95K to a commensurate ordered state where ferromagnetic chains order antiferromagnetically with respect with to their neighbors~\cite{Kobayashi2000,Heid1995}.   Although it is clear that these orders exist out to finite transverse field, the precise phase diagrams is still unknown.   The best information on the phase boundaries thus far comes from heat capacity experiments~\cite{ongcv} that show that at least in the incommensurate ordered phase, which onsets at zero magnetic field at 3 K persists out to a critical field approximately 5.3 T.  The phase boundary of the incomensurate phase is such that one passes through the phase boundary at 2.5K at 2.5 Tesla and at 1.5 K at 4.5 Tesla.  We sketch one scenario for the phase boundaries in Fig.~S\ref{PhaseDiagram}, but the precise extent of of the incommensurate and commensurate phases is unclear.   However, the identification of 1 + 1 D quantum criticality in this work, that of Refs.~\cite{ongcv,imai2014}, and the work of Ref.~\cite{Coldea2010} implies a scenario where there is an effective 2D quantum critical point of an effective 1D system that is found at a field a bit less than the actual quantum critical point associated with the loss of 3D long range order in the real material.  Spectroscopically one finds that the effects of ordering on the THz spectra are weak in the incommensurate phase.   The sharp bound state spectra only appear when one enters the low temperature commensurate state.

\section{Numerical Results}

\begin{figure}[t]
    \centering
    \includegraphics[width=\linewidth]{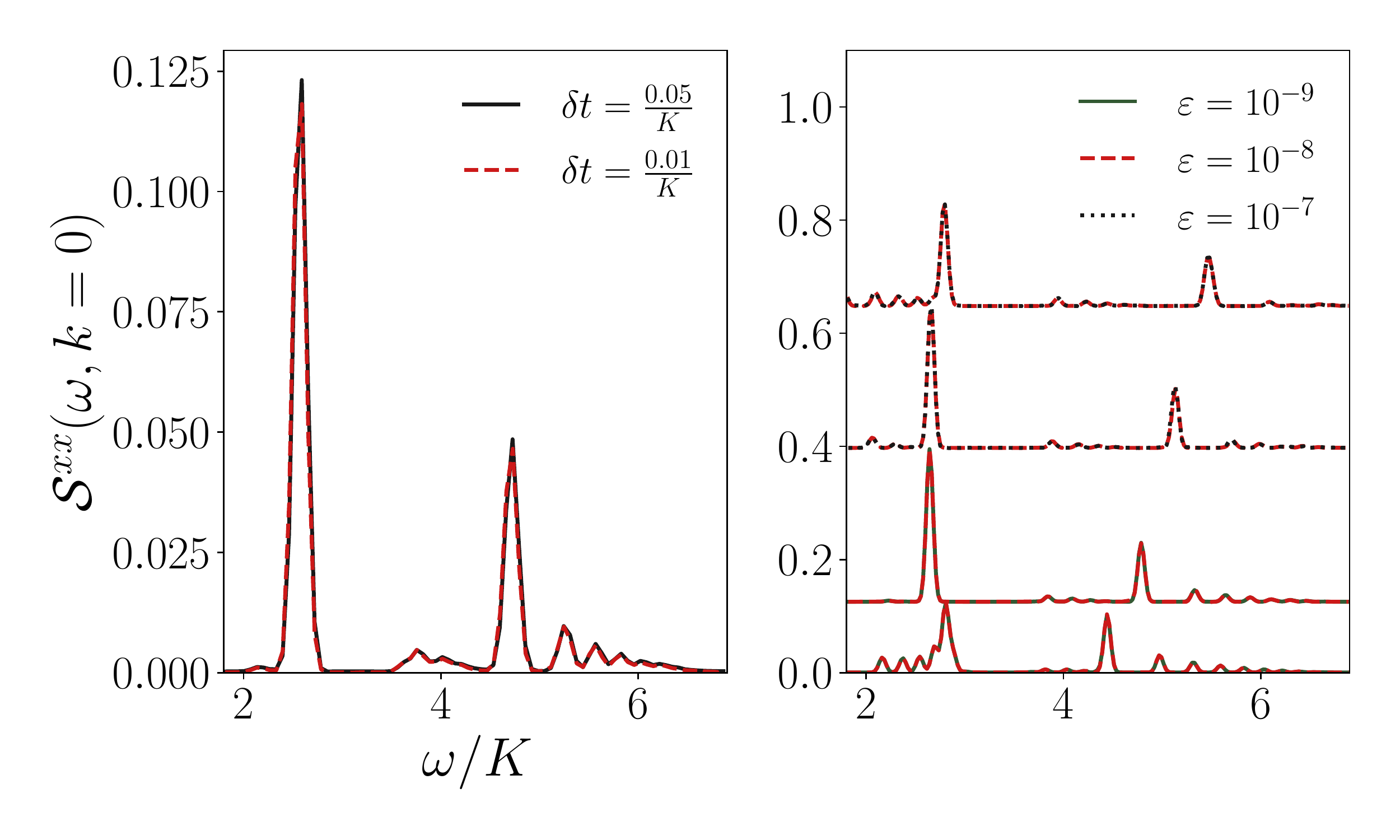}
    \caption{Convergence of the spectral function, $\mathcal{S}^{xx}(\omega,k)$, plotted against frequency at zero momentum with respect to the tDMRG parameters is shown (the energy is measured in units of $K$). The figure on the left shows convergence with respect to the Trotter step, $\delta t$, where during the time evolution simulation, singular values  $ \leq \varepsilon=10^{-8}$ and $ \leq \varepsilon=10^{-9}$ are truncated for $\delta t=\frac{0.05}{K}$ and $\delta t=\frac{0.01}{K}$ respectively. Here, a system of size $L=100$ with $\theta=17^{\circ}$, $h_x=0.3 \, K $ and $h_z=0.035 \, K$ is time evolved for $t_{\text{max}}=\frac{97}{K}$ with $\eta=\frac{20}{t^2_{\text{max}}}$. The peak positions and intensities do not change very much on reducing $\delta t$ from $\frac{0.05}{K}$ to $\frac{0.01}{K}$. On the right is the convergence analysis with respect to the truncation, $\varepsilon$, at $\delta t = \frac{0.05}{K}$. Here $L=512$, $t_{\text{max}}=\frac{290}{K}$ and $\eta=\frac{50}{t^2_{\text{max}}}$ is used. The curves for different $h_x$ have been vertically displaced for clarity: $h_x=0.2 \, K ,0.3 \, K,0.4 \, K,0.5 \, K$ data has been plotted from bottom to top. The curves seem to saturate for $\varepsilon \leq 10^{-8}$ at $h_x =0.2 \, K,0.3 \, K$ and for $\varepsilon \leq 10^{-7}$ at $h_x=0.4 \, K,0.5 \, K$ (the linewidths are larger than the differences in the superimposed curves).}
    \label{fig:tDMRGprm_conv}
\end{figure}

Our Hamiltonian, $\mathcal{H}$, is simulated on an open chain with $L$ sites, using the iTensor C++ library~\cite{itensor}. We compute the structure factor, $\mathcal{S}^{xx}(\omega,k)$, which is the Fourier transform of a two-point correlation in space and time, $S^{xx}(t,x)$, defined as:
\begin{equation}
    S^{xx}(t,r_i-\frac{L}{2})=\langle \tau^{x}_{i} (t) \tau^{x}_{\frac{L}{2}}(0) \rangle.
\end{equation}
The average here is taken in the ground state of $\mathcal{H}$, since we are working at zero temperature. The matrix product state (MPS) approximation of the ground state, $|0\rangle $, is computed using density matrix renormalization group (DMRG). We then compute $| \psi (0) \rangle = \tau^x_{\frac{L}{2}} |0 \rangle $ which is time evolved to get $| \psi(t) \rangle = e^{-i \mathcal{H} t} |\psi(0)\rangle $. Clearly, 
\begin{equation}
    S^{xx}(t,r_i-\frac{L}{2})=e^{i \omega_0 t} \langle 0| \tau^x_i | \psi(t) \rangle
\end{equation}
where $\omega_0$ is the energy of the ground state.

\begin{figure}[t]
  \begin{minipage}[b]{0.9\linewidth}
    \centering
    \includegraphics[width=0.9\linewidth]{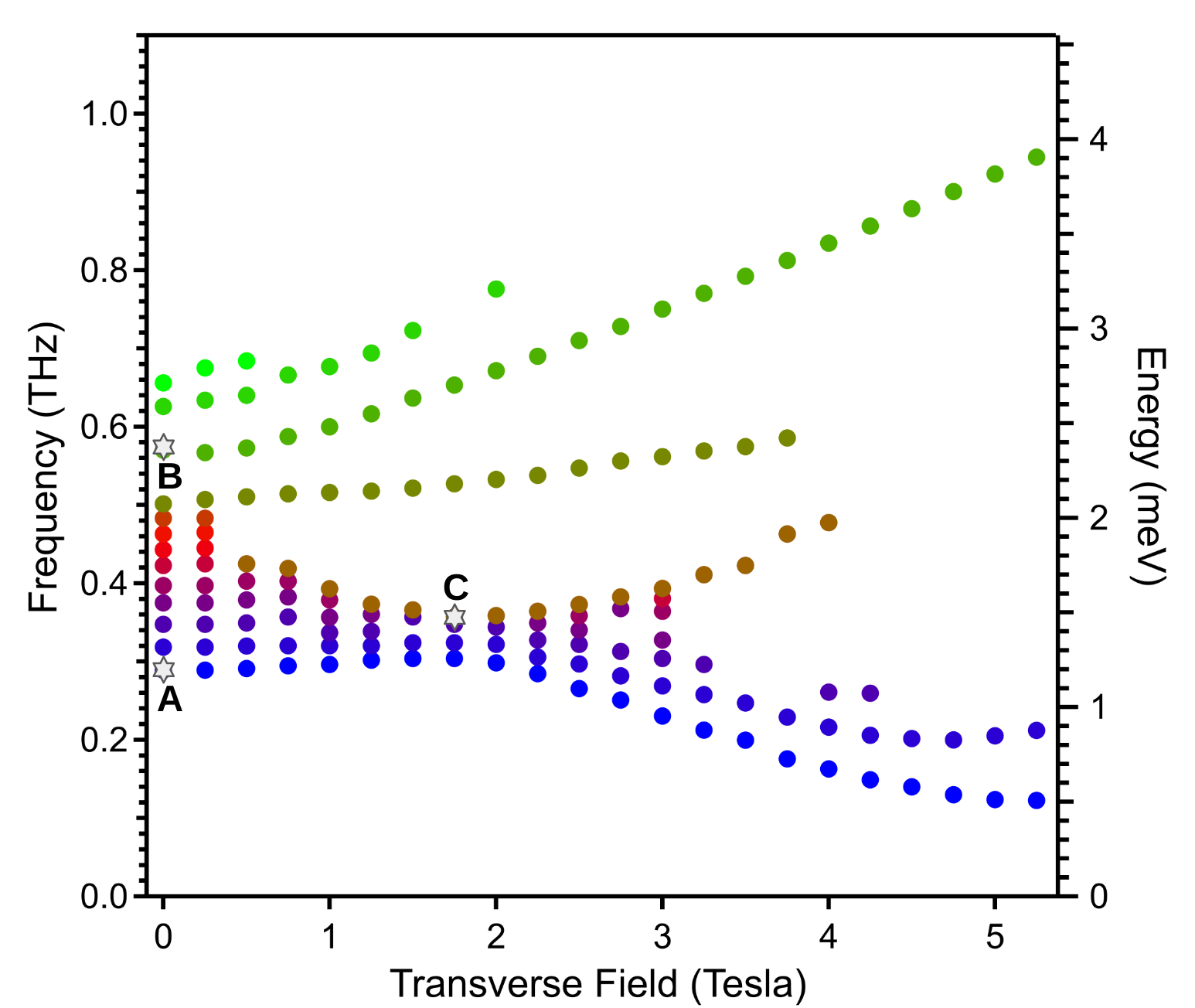}
  \end{minipage}
  
  \vspace*{1cm}
  \begin{minipage}[b]{0.9\linewidth}
    \centering
\begin{tabular}[b]{ |c|c|c| }
  \hline
  \multicolumn{3}{|c|}{Numerical data} \\
  \hline \hline
   $\theta$ (degrees) & $E_B/E_A$ & $E_C/E_A$ \\ \hline
   14  & 1.76 & 1.17 \\ \hline
   15  & 1.83 & 1.17 \\ \hline
   16  & 1.9 & 1.21 \\ \hline
   17  & 2.0 & 1.24 \\ \hline
   18  & 2.07 & 1.28 \\ \hline
   20  & 2.31 & 1.34 \\ \hline
\end{tabular}
\end{minipage}
\caption{Strategy to fit theoretical model parameter $\theta$ to experimental data. Plot of experimental data showing excitation energies as a function of transverse field (energies extracted from Fig 2 in main text). We compare the energies at points \textbf{A}, \textbf{B} and \textbf{C} marked on the experimental plot shown above with corresponding points in our simulated transverse field scans for $\theta=14^{\circ},15^{\circ},16^{\circ},17^{\circ},18^{\circ},20^{\circ}$. The ratios of the energies at points \textbf{B} and \textbf{C}, $E_B$ and $E_C$ respectively, to the energy at point \textbf{A}, $E_A$, are as shown in the above table. Experimentally, $E_B/E_A=1.97$ and $E_C/E_A=1.24$. One can see from the above table, that $\theta=17^{\circ}$ matches the experimental data the best.}
\label{fig:theta_comparison}
\end{figure}

The structure factor is defined as follows:
\begin{equation}
    \mathcal{S}^{xx}(\omega,k)=\int^{\infty}_{-\infty} d t \,  e^{-i \omega t} \sum_i \, e^{i k (r_i-\frac{L}{2})} S^{xx}(t,r_i-\frac{L}{2})
\end{equation}

We compute the structure factor numerically as:

\begin{multline}
    \mathcal{S}^{xx}(\omega,k)=\sum_i e^{i k (r_i-\frac{L}{2})} \, \\  \Re \Bigg[ { \int^{t_{\text{max}}}_{0} d t \,  e^{-i \omega t}  S^{xx}(t,r_i-\frac{L}{2})} \mathcal{W}(t) \Bigg]
\end{multline}

Here we use a Gaussian windowing function, $\mathcal{W}(t) = e^{-\eta t^2}$, in order to avoid artifacts in frequency space when the time data abruptly ends at $t=t_{\text{max}}$. 

\begin{figure}[t]
    \centering
    \includegraphics[width=0.9\linewidth]{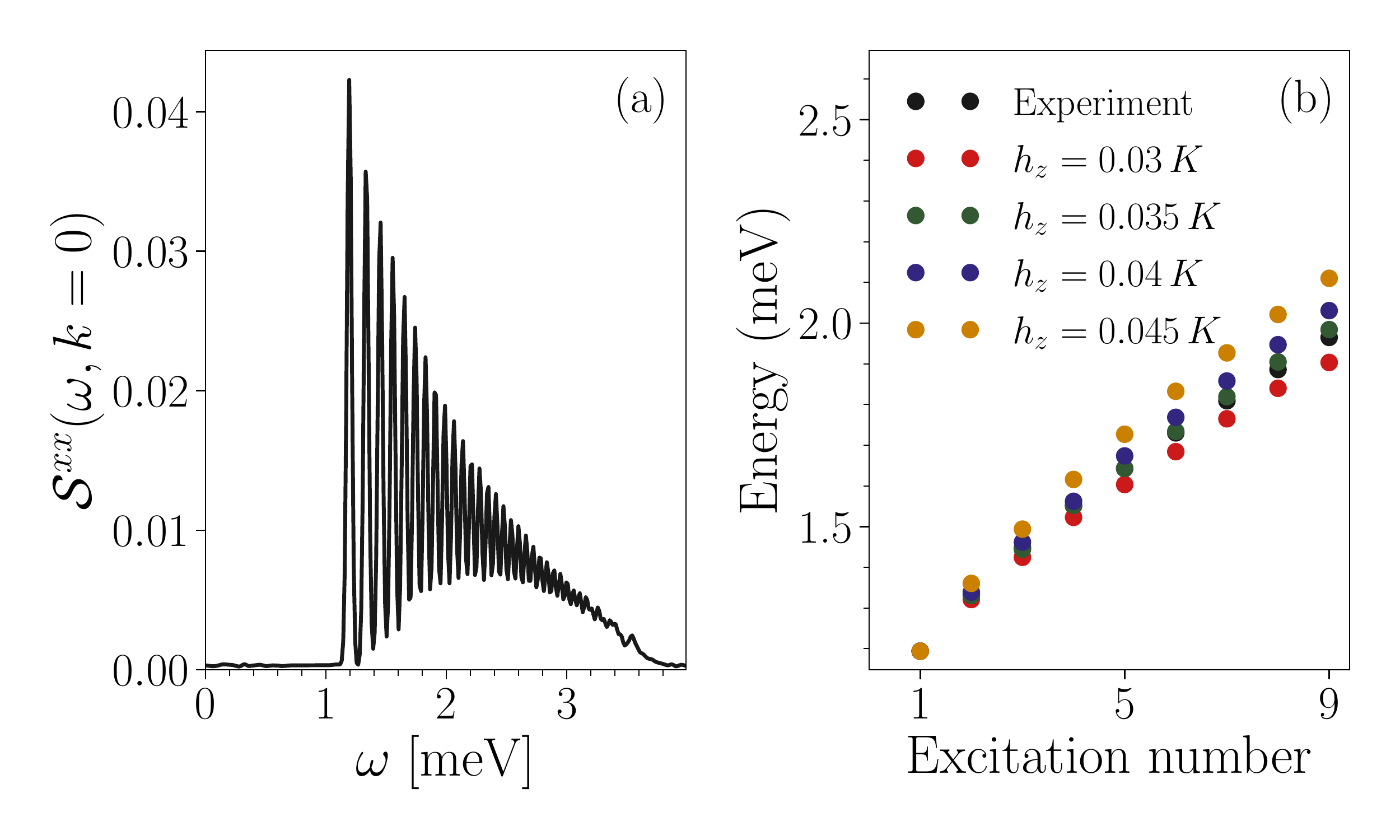}
    \includegraphics[width=0.5\linewidth]{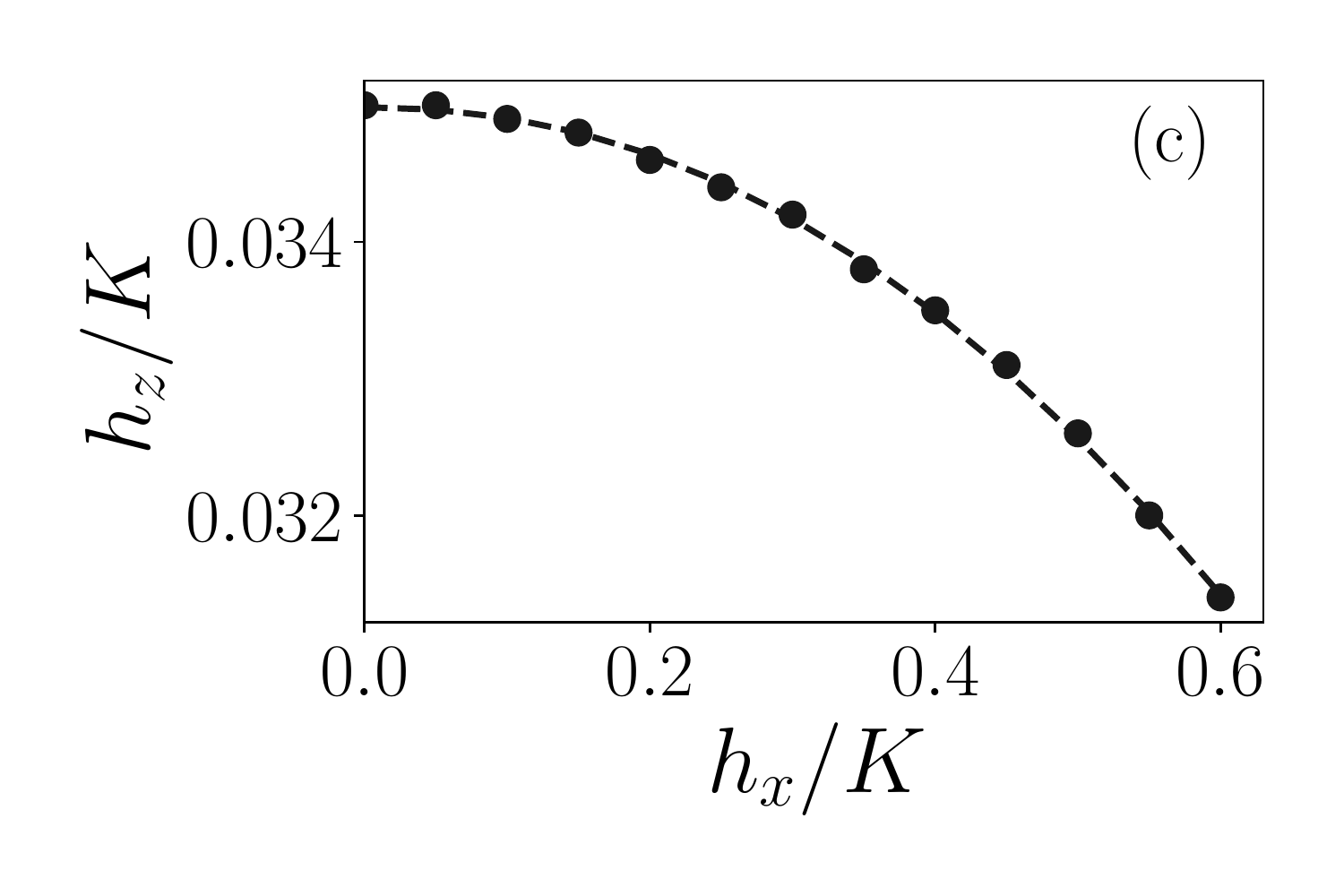}
    \caption{ Determining theoretical parameter $h_z$ from experimental data. (a) The structure factor at zero momentum plotted as a function of frequency, $\omega$, for an $L=512$ system with $t_{\text{max}}=\frac{290}{K}$, $\eta=\frac{50}{t^2_{\text{max}}}$, $\delta t=\frac{0.05}{K}$ and $\varepsilon=10^{-9}$ at $K=0.57\,\text{meV}$, $h_x=0.0$ and $h_z=0.035 \, K$. (b) The excitation energies measured in the experiment are compared with those measured in the tDMRG simulation of an $L=700$ size system for different values of $h_z$ at $h_x=0.0$. $h_z=0.035\,K$ can be seen to match the experiment the best, therefore we use this value of $h_z$ for our simulation. (c) The value of $h_z$ should be determined self consistently from the simulation : $h_z = a \, m_z$ where $a$ is a proportionality constant and $m_z$ is the magnetization of the system. It can be seen that the value of this effective $h_z$ does not vary appreciably in the window we are studying, $0 \leq h_x \leq 0.6 \, K$. Therefore, we keep the longitudinal field a constant, $h_z=0.035 \, K$, in our simulation.}
    \label{fig:excitation_ener}
\end{figure}

\subsection{Convergence analysis}
The time evolution is carried out up to some maximum time $t_{\text{max}}$ using time dependent DMRG (tDMRG) algorithm. There are two sources of errors in the simulation:
\begin{itemize}
    \item \textbf{The Trotter time step}, $\delta t$: The algorithm used has a second order error in the time step. We use $\delta t=0.05/K$ in our simulation. The structure factor at zero momentum is shown to be converged with respect to $\delta t$ in Fig~\ref{fig:tDMRGprm_conv}.
    \item \textbf{The singular value cutoff}, $\varepsilon$: Singular values of the matrix product states below a certain threshold, $\varepsilon$, are truncated during the simulation. We use $\varepsilon=10^{-8}$ and $\varepsilon=10^{-7}$ for transverse fields, $h_x \leq 0.4 K$ and $h_x > 0.4 K$ respectively. The convergence with respect to this parameter is also shown in Fig~\ref{fig:tDMRGprm_conv}.
\end{itemize}

\subsection{Fit to model parameters}

Our model Hamiltonian, $\mathcal{H}$, has the following fit parameters, in addition to the twisted Kitaev coupling $K$ which sets the energy scale ($y$-axis of Fig. 2 in the main paper) and the $g$ sets the scale for the magnetic field ($x$-axis of main paper): 
\begin{itemize}
    \item \textbf{Angle}, $\theta$: We studied excitation energies as a function of $h_x$ (as shown in Fig 2 in the main text) for a few different values of $\theta$. We find that $\theta=17^{\circ}$ matches the experimental data the best as explained in the caption of Fig~\ref{fig:theta_comparison}. 
    \item \textbf{Longitudinal field}, $h_z$: On fixing $\theta=17^{\circ}$, we find that the excitation energies observed in our simulations at $h_x=0$, agree most with those measured in the experiment for $h_z=0.035 \, K$ and $K=0.57\,\text{meV}$  as shown in Fig~\ref{fig:excitation_ener}(b). When $h_x>0$, the effective $h_z$ which is a self-consistent Weiss field should decrease due to the reduced magnetization of the nearest neighbouring chains. However, as shown in Fig~\ref{fig:excitation_ener}(b), this effective $h_z$ does not change very much in the window, $0 \leq h_x \leq 0.6 \, K$ that we study. Hence we choose $h_z=0.035 \, K$ constant for all $h_x$ in our simulations.
\end{itemize}

\begin{figure}[t]
    \centering
    \includegraphics[width=0.6\columnwidth]{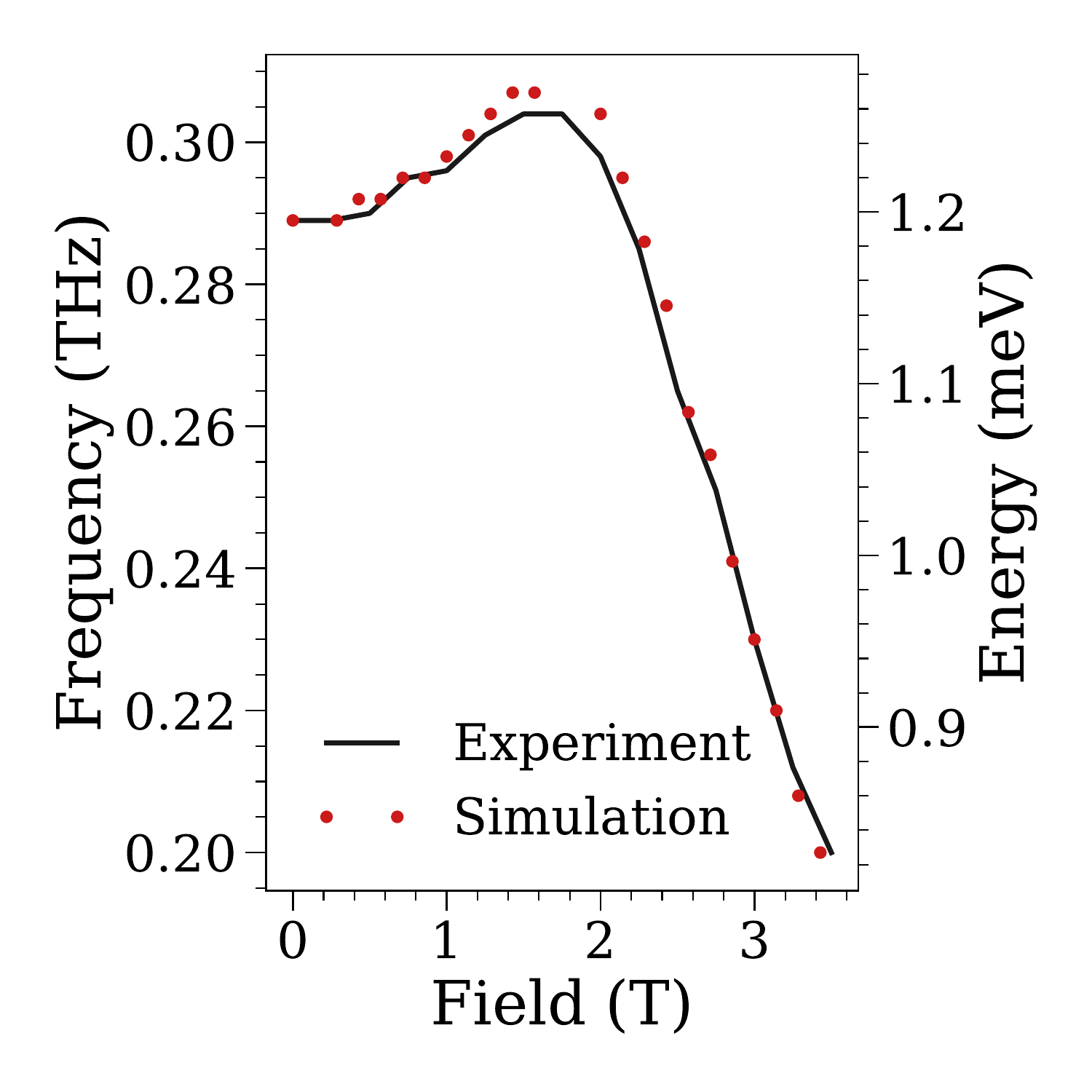}
    \caption{Determining the parameter $g$ that sets the scale of the magnetic field ($x$-axis of Fig 2 in main text). The numerical data was collected in the window $0 \leq h_x \leq 0.6 \, K$ with $\theta=17^{\circ}$, $h_z=0.035\,K$ and $K=0.57\,\text{meV}$. The value of $g$ required to convert $h_x$ from units of energy to magnetic field is determined by requiring that the numerical data matches the experimental data. The figure shows the comparison of the lowest energy band from the experiment and the simulation for $g=3.46$.}
    \label{fig:gfit}
\end{figure}

In order to determine the $g$-factor, $g$, we require that the experimental and numerical data match for the chosen scale of the $x$-axis on Fig 2 of the main text. Fig~\ref{fig:gfit} shows that on choosing $g=3.46$, the energies of the first excited state from the simulation (with $\theta=17^{\circ}$, $h_z=0.035\,K$ and $K=0.57\,\text{meV}$) agree reasonably well with those from the experiment. 

\subsection{Comparison with perturbation theory calculation}

\begin{figure}[h]
    \centering
    \includegraphics[width=0.9\linewidth]{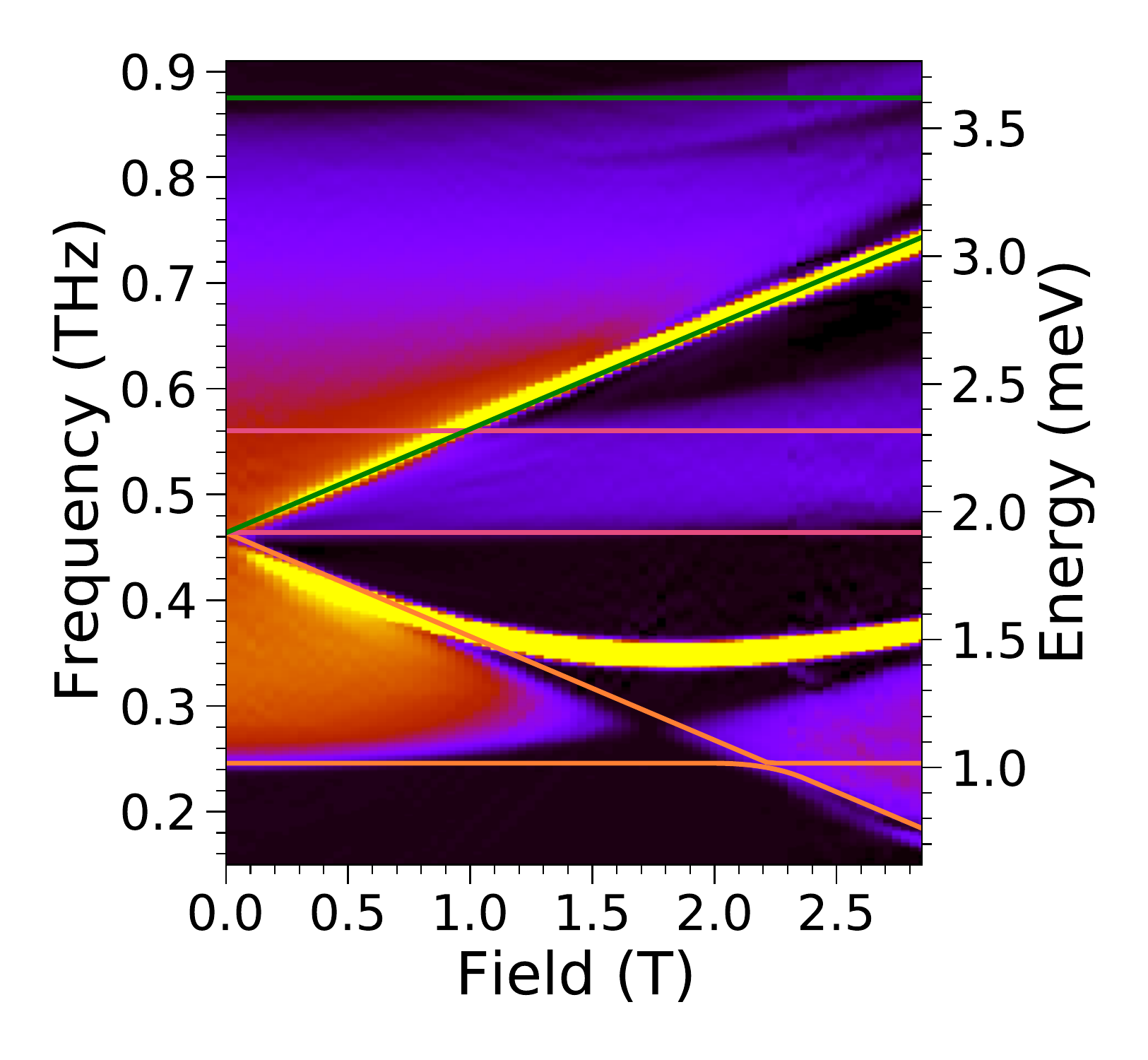}
    \caption{Comparison of the boundaries of the three branches of two domain wall kinetic energy continua calculated in perturbation theory (shown in Fig 1(e) of the main text) with numerical simulation data (shown in Fig 2(d) of main text) for $\theta=17^{\circ}$, and $K=0.57\,\text{meV}$.}
    \label{fig:pert_comparison}
\end{figure}

In Fig.~\ref{fig:pert_comparison} we show the same DMRG spectral function data as in the main section of the manuscript and with the boundaries of the three branches of the two domain wall kinetic energy that arise from perturbation theory. This comparison has no fit parameters. We find reasonable qualitative agreement validating the physical interpretation of the three regions as arising from the band structure of the domains walls. Corrections to the continua arise from interaction between domain walls and higher domain wall states, which we have ignored in the crude perturbative calculation.

\end{document}